%% file: main.tex
\documentclass{eptcs}
\providecommand{\event}{QPL 2012} 
\usepackage{breakurl}             
\usepackage{amssymb}
\usepackage{amsmath}
\usepackage{stmaryrd}
\usepackage{prooftree}
\usepackage{paralist}
\usepackage{url}
\usepackage{datetime}
\usepackage{tikz}
\usetikzlibrary{trees,arrows,positioning,calc,decorations.pathmorphing}
\usepackage[all]{xy}

\input{macros}

\input{Qcircuit}

\title{Application of Quantum Process Calculus to Higher Dimensional Quantum Protocols}
\author{Simon J. Gay
\institute{School of Computing Science\\
University of Glasgow, UK}
\email{Simon.Gay@glasgow.ac.uk}
\and
Ittoop Vergheese Puthoor \thanks{Supported by a Lord Kelvin / Adam Smith Scholarship from the University of Glasgow.}
\institute{School of Computing Science and\\
School of Physics and Astronomy\\
University of Glasgow, UK}
\email{ittoop@dcs.gla.ac.uk}
}
\def\titlerunning{Application of Quantum Process Calculus to Higher Dimensional Quantum Protocols}
\def\authorrunning{S. J. Gay and  I. V. Puthoor}
\begin{document}
\maketitle


\begin{abstract}
We describe the use of quantum process calculus to describe and
analyze quantum communication protocols, following the successful
field of \emph{formal methods} from classical computer science. We 
have extended the quantum process calculus to describe $d$-dimensional
quantum systems, which has not been done before. We summarise the 
necessary theory in the generalisation of quantum gates and Bell states
and use the theory to apply the quantum process calculus CQP to quantum
protocols, namely qudit teleportation and superdense coding.

\end{abstract}

\input{introduction}

\input{Theory}

\input{CQP}

\input{SDC}

\input{conclusion}

\bibliographystyle{eptcs}
\bibliography{report_main}

\end{document}

%% file: macros.tex
\newtheorem{definition}{Definition}

\newcommand{\mktype}[1]{\mathsf{#1}}

\newcommand{\Qdit}{\mktype{Qdit}}

\newcommand{\Val}{\mktype{Val}}

\newcommand{\mkterm}[1]{\mathsf{#1}}

\newcommand{\unit}{\mkterm{unit}}

\newcommand{\Prob}[1]{\boxplus_{#1}}
\newcommand{\ms}[1]{|{#1}|^{2}}
\newcommand{\cnfig}[3]{({#1};{#2};{#3})}

\newcommand{\ptrns}[3]{{#1}\stackrel{#2}{\longrightarrow}{#3}}

\newcommand{\mkRrule}[1]{\mbox{\textsc{R-#1}}}

\newcommand{\Rplus}{\mkRrule{Plus}}
\newcommand{\Rmeasure}{\mkRrule{Measure}}
\newcommand{\Rcontext}{\mkRrule{Context}}
\newcommand{\Rtrans}{\mkRrule{Trans}}

\newcommand{\typed}[2]{{#1}\mathrel{\!:\!}{#2}}

\newcommand{\ket}[1]{|#1\rangle}
\newcommand{\nil}{\mathbf{0}}
\renewcommand{\parallel}{\mathbin{\mid}}
\newcommand{\inp}[2]{{#1}?{[#2]}}
\newcommand{\outp}[2]{{#1}!{[#2]}}

\renewcommand{\vec}[1]{\widetilde{#1}}
\newcommand{\new}{\mathsf{new}\ }

\newcommand{\qdit}{\mathsf{qdit}\ }
\newcommand{\bit}{\mathsf{bit}}

\newcommand{\chant}[1]{\widehat{~}[#1]}

\newcommand{\qgate}[1]{\mathsf{#1}}
\newcommand{\pname}[1]{\mathit{#1}}
\newcommand{\action}[1]{\{#1\}}
\newcommand{\trans}{\mathbin{*\!\!=}}
\newcommand{\sep}{\,.\,}

\newcommand{\measure}{\mathsf{measure}\ }

\newcommand{\qstore}[2]{[#1 \mapsto #2]}

\newcommand{\transition}[1]{\stackrel{#1}{\longrightarrow}}
\newcommand{\transitione}{\longrightarrow_e}
\newcommand{\transitionv}{\longrightarrow_v}

\newcommand{\ptrans}[1]{\stackrel{#1}\rightsquigarrow}
\newcommand{\weaktrans}[1]{\stackrel{#1}{\Longrightarrow}}

\newcommand{\cfgdistsum}{\oplus}

\newcommand{\Dist}[2]{\cfgdistsum_{#1}~#2~}

\newtheorem{example}{Example}

\newcommand{\subst}[2]{\{{#1}/{#2}\}}

\newcommand{\bra}[1]{\langle#1|}

\newcommand{\parcomp}{~\Vert~}

\newcommand{\gX}{\qgate{X}}

\newcommand{\gZ}{\qgate{Z}}
\newcommand{\gH}{\qgate{H}}
\newcommand{\gRC}{\qgate{R}_{c}}
\newcommand{\gLC}{\qgate{L}_{c}}

\newcommand{\ltrm}[3]{\lambda{#1}\bullet{#2}; {#3}}
\newcommand{\ltrmshort}[2]{\lambda{#1}\bullet{#2}}

\renewcommand{\vec}[1]{\widetilde{#1}}

\newcommand{\pteleport}{\mathit{Teleport}}

\newcommand{\palice}{\pname{Alice}}
\newcommand{\pbob}{\pname{Bob}}

\newcommand{\psdc}{\pname{SDC}}

%% file: introduction.tex
\section{Introduction}
\label{sec-intro}

Quantum computing and quantum communication have attracted great interest as quantum computing offers great improvements in algorithmic efficiency and quantum cryptography helps to provide more secure communication systems.  Quantum computing, with its inherent parallelism from the superposition principle of quantum mechanics, offers the prospect of vast improvement over classical computing. The most dramatic result is that Shor \cite{Shor1994} showed a quantum algorithm which is more efficient than any known classical algorithm for factorisation of integers. 

Quantum process calculus is a particular field of formal languages which is used to describe and analyse the behaviour of systems that combine both quantum and classical computation and communication. \emph{Formal methods} provide theories and tools which can be used to specify, develop and verify systems in a systematic manner. This field has been successful in classical computer science and to use these mathematically based techniques to describe quantum systems is one reason for developing quantum formal methods. Another motivation is that it supports the development of automated tools which can be used in verifying the correctness of practical quantum  technologies such as quantum cryptographic systems.

Our own approach is based on a particular quantum process calculus
called Communicating Quantum Processes (CQP), developed by Gay and
Nagarajan \cite{Gay2005}. Recent work on CQP has shown that the idea of \emph{behavioural equivalence} between processes  is a congruence, meaning that it is preserved by
inclusion in any environment. This has been reported in Davidson's Ph.D thesis
\cite{DavidsonThesis}. The aim is to prove the correctness of a system by using the following methodology. Initially, define a \emph{System} which is a process that models the system of interest. Then define the \emph{Specification}, a simpler process that exhibits the desired behaviour of the \emph{System}. Finally, we prove that these two processes are equivalent. This approach has been illustrated by analysing quantum teleportation, superdense coding and a quantum error correction system. A similar theory has also been developed independently for qCCS by Feng \emph{et al.}\cite{Feng2011}.

The quantum process calculi which have been developed to date are defined for modelling systems that involves qubits, which are transmitted from process to process along communication channels. Experiments in quantum optics show that the physical systems that represent quantum information processing need not be limited to quantum bits (qubits) but can use higher dimensional systems, i.e. qudits (a quantum system with $d$-dimensional Hilbert space) \cite{Dada2011}. A photon can carry both spin and orbital angular momentum (OAM) and either or both of these properties can be used to represent quantum information. The spin angular momentum is associated with polarisation of light and is described completely within a two dimensional Hilbert space. But the OAM states of light constitute an infinite-dimensional Hilbert space with orthonormal basis states $\ket{l}$ characterised by an azimuthal phase factor $exp(il\phi)$, carrying an OAM of $l\hslash$ per photon,
 \begin{equation}
 \ket{\psi} = \sum^{\infty}_{l=-\infty}a_{l}\ket{l}
 \end{equation}
Restricting to a finite number of basis states then leads to the implementation of qudits, which carry quantum information in a finite $d$-dimensional basis. Experiments have shown that photon pair entangled in their OAM up to $|l| = 20$, can be produced with high-fidelity by a parametric down-conversion process \cite{Jack2009}.  In relation to quantum computation and communication, the higher dimensional Hilbert space of orbital angular momentum allows the implementation of new quantum protocols, which can offer higher information capacity and greater degree of security \cite{Arnold2008}. Recent studies have adopted the higher dimensionality encoded in the polarisation and orbital angular momentum for quantum information and cryptographic processing \cite{Ambrosio2012}. Because of growing interest in the use of higher-dimensional systems for quantum information processing, we want to extend the theory of quantum process calculus to higher dimensions. 

The quantum teleportation protocol plays a vital role in quantum information theory. The concept was first proposed by Bennett \emph{et al.}\cite{Bennett1993} and has been an active study of research.  This is a process by which a qubit can be transmitted exactly from one location to another, without communicating it through the intervening space. In qudit teleportation, we consider an arbitrary qudit state to be teleported  with the help of  maximally entangled states or the so-called \emph{Bell states} of two other qudits. Teleportation is a standard protocol and we have quantum circuits which describe the protocol. But unlike the circuit model, the CQP model of the protocol is able to clearly describe the actions of the processes involved. 

The theory of equivalence have been developed in CQP for qubits and not qudits. Our future task would be to extend the theory of equivalence to qudits and in this paper, we establish the foundation of using equivalence in higher-dimensions by providing the definitions of CQP for qudits. With the help of these definitions, we analyse the quantum protocols namely qudit teleportation and superdense coding. 

The rest of the paper is organised as follows. In Section~\ref{sec:THEORY} we provide the basic
definition of qudit and the generalisation of the quantum operators for \emph{d} 
dimensional systems. We review the language of CQP in Section ~\ref{sec:CQP} and illustrate it with a model of a qudit teleportation. Section ~\ref{sec:SDC} provides the description and execution of the superdense coding protocol in higher dimensions. Section ~\ref{sec:conclusion} concludes with an indication of directions of future work.

\paragraph*{Related Work}

All the quantum process calculi which have been established so far, involved qubits and not qudits. Lalire and Jorrand developed the quantum process calculus called Quantum Process Algebra (QPAlg)\cite{Lalire2004} and Feng \emph{et al.} \cite{Feng2006} developed qCCS, a quantum extension of the classical value-passing CCS \cite{Milner1989}. The theory of equivalence has been defined for both QPAlg \cite{Lalire2006} and qCCS  \cite{Feng2011}. This is the first time that the quantum process calculus (CQP) have been extended to higher dimensions.



%% file: Theory.tex
\section{Quantum Gates for $d$-dimensional Systems}
\label{sec-THEORY}
\label{sec:THEORY}

 Any physical system is associated with a \emph{Hilbert space}, called its \emph{state space}. We denote by $\mathbb{H}$, the Hilbert space for a qudit, that is a $d$-dimensional vector space over the complex numbers, $\mathbb{C}$ with a basis denoted by $\{\ket{0},\ket{1},\sep\sep\sep,\ket{d-1}\}$. We fix each orthonormal basis state of the $d$-dimensional Hilbert space to correspond to an element of $\mathbb{Z}_{d}$; as such the basis $\{\ket{0},\ket{1},\sep\sep\sep,\ket{d-1}\}\subset  \mathbb{C}^{d}$ is called the \emph{computational basis} \cite{Bartlett2002,Nikolopoulos2005}. The system is completely described by a unit vector $\ket{\psi}$ within its state space, called the \emph{state vector}. Our main system of interest is the \emph{qudit}, a $d$-dimensional quantum state $\ket{\psi} \in \mathbb{H}$. The set of vectors $\{\ket{0},\ket{1},\sep\sep\sep,\ket{d-1}\}$ is also called the \emph{standard basis} of the qudit state space $\mathbb{C}^{d}$. An $n$-qudit state, in the tensor product Hilbert space is given by $\mathbb{H}^{\otimes n}= (\mathbb{C}^d)^{\otimes n} = \mathbb{C}^d \otimes \sep\sep\sep \otimes \mathbb{C}^d $. The \emph{standard basis}  of $\mathbb{H}^{\otimes n}$ is the orthonormal basis given by the $d^{n}$ classical $n$-qudits.
 \begin{equation}
 \ket{i_{1}i_{2}\sep\sep\sep i_{n}} = \ket{i_{1}}\otimes \ket{i_{2}}\otimes\sep\sep\sep\otimes\ket{i_{n}}
 \end{equation}
where 0 $\le \emph{i}_{j} \le \emph{d-1}$.
We can write the general state of a qudit as 
 \begin{equation}
 \ket{\psi} = \sum^{d-1}_{i=0}\alpha_{i}\ket{i}
 \end{equation}
where $\alpha_{i} \in \mathbb{C}$ and $\Sigma^{d-1}_{i=0} \mid\alpha_{i}\mid^{2} = 1$. 

\subsection{Generalised CNOT Gate}

We now introduce the elementary quantum gates or operators for $d$-dimensional systems.
Let $\mathbb{H_{A}}$ and $\mathbb{H_{B}}$ be $d$-dimensional Hilbert spaces, consider the set of $d^{2} \times d^{2}$ unitary transformations $U \in U(d^{2})$ that act on the two-qudit quantum system $\mathbb{H_{A}} \otimes \mathbb{H_{B}}$. To generalise the NOT and the CNOT gates, we note that in the context of qubits, the NOT gate, is basically a mod-$2$ adder. For qudits this operator gives way to a mod-$d$ adder, or a CNOT Right-Shift gate. Let $R_{C} \in U(d^{2})$ represent the generalised CNOT Right-Shift gate that has control qudit $\ket{\psi} \in \mathbb{H_{A}}$  and target qudit $\ket{\phi} \in \mathbb{H_{B}}$. The action of $R_{C}$ on the set of standard basis states $\ket{m} \otimes \ket{n}$ of $\mathbb{H_{A}} \otimes \mathbb{H_{B}}$ is given by
 \begin{equation}
R_{C}\ket{m} \otimes \ket{n} = \ket{m} \otimes \ket{n \oplus m},\hspace{4cm} \emph{m,n} \in \mathbb{Z}_{d}
 \end{equation}
with $\oplus$ denoting addition modulo $d$. Similarly, $L_{C} \in {U} (d^{2})$ denotes the generalised CNOT Left-Shift gate.
 \begin{equation}
R^{-1}_{C}\ket{m} \otimes \ket{n} \equiv \emph{L}_{C}\ket{m} \otimes \ket{n} = \ket{m} \otimes \ket{n \ominus m}\hspace{4cm}
 \end{equation}
 
 \subsection{Generalised Pauli Gates}
 
 The next set of operators which are used to perform theoretical investigations of quantum systems are the Pauli operators. We now define the generalised Pauli operators for $d$-level quantum systems \cite{Bartlett2002}.
  \begin{equation}
  {U} = \{X^{j}Z^{k}: {j,k} \in \mathbb{Z}_{d}\}.
 \end{equation}
 where $X$ and $Z$ are defined by their action on the standard basis
 \begin{equation}
  X^{j}\ket{m} = \ket{m \oplus j},
 \end{equation}
 \begin{equation}
  Z^{k}\ket{m} = exp(2\pi ikm/d) \ket{m} = \omega^{km}\ket{m},
 \end{equation}
 where 
 \begin{equation}
 \omega \equiv exp(2\pi i/d) .
 \end{equation}
The indices  $j$ and $k$ refer to shift and phase changes in the standard basis, respectively. Therefore the generalised Pauli operators can be represented in the form
 \begin{equation}
 {U}_{jk} = \Sigma_{m \in \mathbb{Z}_{d}}\omega^{km}\ket{m \oplus j}\bra{m}
\end{equation}
Note that $X$ and $Z$ do not commute; they obey
  \begin{equation}
  Z^{k}X^{j} = \omega^{jk}X^{j}Z^{k}
 \end{equation}
 and $X^{d} = Z^{d} = I$. 
 
 \subsection{Generalised Hadamard Gate and Bell States}
 
We now define a generalisation of the Hadamard gate which is useful in manipulating qudits for various applications \cite{Fujii2001}.
\begin{equation}
 H\ket{j} =  \frac{1}{\sqrt{d}}\sum_{m=0}^{d-1}\omega^{-jm}\ket{m}
\end{equation}
This operator is also known as the quantum Fourier transform when $d = 2^{n}$. In that case it acts on $n$ qudits. Here we assume it to be a basic gate on one single qudit, in the same way that the ordinary Hadamard gate is a basic gate on one qubit. This operator is symmetric and unitary, but not Hermitian.
 
 A generalisation of the familiar Bell states for qudits has been introduced in \cite{Barnett2009}. The entangled state $\ket{\Psi^{nm}}_{AB}$ is called the \emph{generalised Bell state} whereby $A$ and $B$ each possess one qudit of this two qudit state. These are a set of $d ^ {2}$ maximally entangled states and can be explicitly written as:
 \begin{equation}
 \ket{\Psi^{nm}} _{AB}= \frac{1}{\sqrt{d}}\sum_{j=0}^{d-1}\omega^{-jn}\ket{j}_{A} \otimes \ket{j \oplus m}_{B}
 \end{equation} 
 where $m$ and $n$ run from $0$ to $d-1$. These states have the properties $\bra{\Psi^{nm}}\ket{\Psi^{n'm'}} = \delta_{nn'} \delta_{mm'}$(orthonormality) and $\emph{tr}(\ket{\Psi^{nm}}\bra{\Psi^{nm}}) = \frac{1}{d}I$(maximal entanglement). Later, we will use the particular Bell state
\begin{equation}
\ket{\Psi^{00}}_{AB} = \frac{1}{\sqrt{d}}\sum_{j=0}^{d-1}\ket{j}_{A} \otimes \ket{j}_{B}
\end{equation}  
To construct the generalised Bell state, we fist apply the Hadamard transform $(H\otimes I)$ to the qudit $A$. This acts on basis states $\ket{n}_{A}\ket{m}_{B}$ as follows 
\begin{equation}
(\emph{H}\otimes\emph{I})\ket{n}_{A}\ket{m}_{B} = \frac{1}{\sqrt{d}}\sum_{j=0}^{d-1}\omega^{-jn}\ket{j}_{A} \otimes\ket{m}_{B}
\end{equation}
where $\omega$ is a primitive  $d^{th}$ root of unity in $\mathbb{C}$ such that $\omega^{d} = 1$. Then we apply CNOT Right-Shift gate after the Hadamard transform and we obtain the generalised Bell state
\begin{equation}
\ket{\Psi^{nm}} _{AB} = \emph{R}_{C}[(\emph{H}\otimes\emph{I})\ket{n}_{A}\ket{m}_{B}] = \frac{1}{\sqrt{d}}\sum_{j=0}^{d-1}\omega^{-jn}\ket{j}_{A} \otimes\ket{ j \oplus m}_{B}
\end{equation}

%% file: CQP.tex
\section{Communicating Quantum Processes (CQP)}
\label{sec:CQP}
\label{sec-CQP}

CQP is a quantum process calculus which was developed by Gay and Nagarajan \cite{Gay2005}. This is used for formally defining the structure and behaviour of systems that are a combination of both quantum and classical communication and computation.  CQP is based on the $\pi$-calculus \cite{Milner1999,Milner1992} with primitives for quantum information. The general idea is that a system is considered to be made up of independent components or \emph{processes}. The \emph{processes} can communicate by sending data along \emph{channels} and these data are qubits or classical values. The earlier operational semantics of CQP are defined with respect to qubits and the full details can be found in Davidson's Ph.D thesis  \cite{DavidsonThesis}. The operational semantics of CQP are defined using reductions under the assumption that quantum systems are closed to any environment. The most distinctive feature of CQP is the inclusion of a static type system, the purpose of which is to classify classical and quantum data and also to enforce the no-cloning property of quantum information. A full treatment of the type system with associated proofs of soundness and a type checking algorithm is presented by Gay and Nagarajan \cite{Gay2006a}. The language has been presented as a solid framework with the ability to easily add new functionality as required.  

We have extended CQP to describe \emph{d}-dimensional quantum systems and in this paper we present the application of CQP in describing two quantum protocols namely teleportation and superdense coding for higher dimensional systems.  Although typing is important, we will not discuss it in detail in the present paper; however, our CQP definitions will include type information because it usually forms useful documentation. We will not be giving the complete formal definition of the language, but will explain it informally in relation to our model of qudit teleportation.

\subsection{Qudit Teleportation}

\begin{figure}
 \Qcircuit @C=1em @R = 2.0em {
    & & & & & & & & & & &\lstick{x = \ket{\psi}}       &     \qw  &  \qw  &  \qw  & \qw & \ctrl{1} & \qw & \gate{H} & \qw & \qw & \qw & \meter & \control \cw \cwx[2]  &\\ 
    & & & & & & & & & & &\lstick{z = \ket{0}}       &     \gate{H} & \ctrl{1} & \qw  & \qw & \gate{L_C} & \qw & \qw & \qw & \meter & \control \cw \cwx[1] &\\   
     & & & & & & & & & & &\lstick{y = \ket{0}}      &     \qw &  \gate{R_C} & \qw & \qw & \qw & \qw & \qw & \qw & \qw &  \gate{X^{-M_1}} & \qw & \gate{Z^{M_2}} & \qw  & \rstick{\ket{\psi}}
}
\caption{\label{fig:Teleportation}Qudit Teleportation}
\end{figure} 

Quantum teleportation \cite{Barnett2009, Fujii2001} is a protocol, which allows two users who share an entangled pair of qudits, to exchange an unknown quantum state by communicating only two classical values depending on the dimension $d$ of the system. The quantum circuit model of the protocol for qudits is shown in Figure~\ref{fig:Teleportation}. This circuit model is similar to the quantum teleportation for qubits. The difference is that we have to use the generalised quantum gates (CNOT and Hadamard) for qudits which we have explained earlier. Although the circuit model represents the teleportation protocol, it defines the operation involved in the protocol, but it does not give a full description of the protocol itself. For example, the circuit model does not explain to us that the protocol consists of a definition of two users and the way in which they communicate, as well as the definition of the quantum operation involved in the protocol. The benefit of using our CQP model is that it not only provides the definition of the system but also be able to give a clear and formal description of the actions of the two users involved in the protocol.

Our model of qudit teleportation protocol consists of two processes:  $\palice$ and $\pbob$,  we say the sender is $\palice$ and the receiver is $\pbob$.  $\palice$ possesses the qudit labelled $x$ which is in some unknown state $\ket{\psi}$; this is the qudit to be teleported. Qudits $y$ and $z$ are an entangled pair, which is generated by applying a Hadamard and CNOT- Right Shift gate to the qudits. The entangled state $\ket{\Psi^{00}}_{zy}$ is given by equation (14). Then qudit $z$ is given to $\palice$ and qudit $y$ is given to $\pbob$.The CQP definition of  $\palice$ is as follows
\[
\begin{array}{ll}
\multicolumn{2}{l}{\palice(\typed{c}{\chant{\Qdit}},\typed{e}{\chant{\Val,\Val}}) = \inp{c}{\typed{x}{\Qdit}}\sep\action{x,z\trans\gLC}\sep\action{x\trans\gH}\sep\outp{e}{\measure z,\measure x}\sep\nil}
\end{array}
\]
$\palice$ is parameterized by two channels, $c$ and $e$. She receives the qudit on channel $c$. The type of $c$ is $\chant{\Qdit}$, which is the type of a channel on which the message communicated is a qudit. Channel $e$ is where $\palice$ sends the classical values resulting from her measurement. Each message on $e$ consists of two classical values, as indicated by the type $\chant{\Val,\Val}$.

The right hand side of the definition specifies the behaviour of $\palice$. The first term, $\inp{c}{\typed{x}{\Qdit}}$ specifies that a qudit is received from channel $c$ and given the local name $x$. Then follows a sequence of terms separated by dots which are an indication of temporal sequencing, from left to right. The term $\action{x,z\trans\gLC}$ specifies that the CNOT- Left Shift operation is applied to qudits $x$ and $z$ and next term $\action{x\trans\gH}$ specifies that the Hadamard operation is applied to qudit $x$. The final term $\outp{e}{\measure z,\measure x}$ indicates that the qudits $x$ and $z$ are measured which results in two classical values ($M_{1}$ and $M_{2}$) ranging from $0$ to $d-1$ (where $d$ is the dimension of the system). These two values are sent as a message on channel $e$. The term $\nil$ simply indicates termination.

We model the process $\pbob$, which receives the two classical values from  channel $e$ (connected  to $\palice$) and outputs the teleported qudit through channel $d$. 
\[
\begin{array}{ll}
 \multicolumn{2}{l}{\pbob(\typed{e}{\chant{\Val,\Val}},\typed{d}{\chant{\Qdit}}) = \inp{e}{\typed{M_{1}}{\Val},\typed{M_{2}}{\Val}}\sep\action{y\trans\gX^{-M_{1}}}\sep\action{y\trans\gZ^{M_{2}}}\sep\outp{d}{y}\sep\nil}
\end{array}
\]
Using the classical values, $\pbob$ performs the necessary unitary operations on his qudit $y$ as indicated by the terms $\action{y\trans\gX^{-M_{1}}}$ and $\action{y\trans\gZ^{M_{2}}}$. By doing this, he can recover the original state $\ket{\psi}$.
The complete system is defined as follows.
\[
\begin{array}{ll}
\multicolumn{2}{l}{\pteleport = (\qdit y,z)(\action{z\trans\gH}\sep\action{z,y\trans\gRC}\sep(\new e)(\palice(c,e) \parallel \pbob(e,d)))}
\end{array}
\]
The complete $\pteleport$ process consists of $\palice$ and $\pbob$ in parallel, indicated by the vertical bar.  Putting the two processes in parallel means that the output on $e$ in $\palice$ synchronises with the input on $e$ in $\pbob$. Channel $e$ is designated as a private local channel. This is specified by ($\new e$), which is a construct from pi-calculus to dynamically create fresh channels. The first term,  $(\qdit y,z)$ in $\pteleport$, allocates two fresh qudits, each in state $\ket{0}$, and gives them the local names $y$ and $z$. The next two terms create an entangled pair with qudits $y$ and $z$.

\subsection{Semantics}

In the previous section, we have described informally the behaviour of the processes in the qudit teleportation system. The precise behaviour can be specified by using the formal semantics of CQP. In this section we will explain the formal semantics, although without giving all of the definitions.

The semantics are defined by labelled transitions between process terms in the same way as in classical process calculus. A transition is labelled by an action which is written as $\transition{\alpha}$ where $\alpha$ is an action. The actions can be classified as  $input, output$ and $internal$ action where $\inp{c}{x}, \outp{c}{x}$ and $\tau$ for input on channel $c$, output on channel $c$, and internal action are used as the respective notations. For example, a process of the form $\outp{c}{x}\sep P$, where $P$ is some continuation process, has the transition
\begin{equation}\label{eq:trans-out}
\ptrns{\outp{c}{x}\sep P}{\outp{c}{x}}{P}.
\end{equation}
If there is another process $Q$ in a system which is ready to receive on channel $c$, then this would become an actual step in the behaviour of a system. The labelled transition representing the potential input is
\begin{equation}\label{eq:trans-in}
\ptrns{\inp{c}{x}\sep Q}{\inp{c}{x}}Q.
\end{equation}
When we consider the two process to be parallel with each other, the input and output actions combine, resulting in a $\tau$ transition which represents a single step of
behaviour:
\[
\ptrns{\outp{c}{x}\sep P \parallel \inp{c}{x}\sep
  Q}{\tau}{P \parallel Q}.
\]

The complete definition of the semantics takes the form of a
collection of labelled transition rules. Transition
(\ref{eq:trans-out}) becomes a general rule for output  and transition (\ref{eq:trans-in}) is a
general rule for input. The interaction between input and output is
defined by the rule
\[
\begin{prooftree}
\ptrns{P}{\outp{c}{v}}{P'} \qquad \ptrns{Q}{\inp{c}{v}}{Q'}
\justifies
\ptrns{P \parallel Q}{\tau}{P' \parallel Q'}
\end{prooftree}
\]
which specifies that if the transitions above the line (hypotheses)
are possible then so is the transition below the line (conclusion). \cite{Milner1999,Sangiorgi2001}.

We need to include a representation of the quantum state in order to define the semantics of CQP. The execution of a system is not fully described by a process term, but also depends on the quantum state. For this reason, the operational semantics are defined using \emph{configurations}, which represent both the quantum state and the process term.  A \emph{configuration} is a tuple ($\sigma;\omega;P$) where $\sigma$ is a mapping from qudit names to the quantum state, $\omega$ is a list of qudit names, and P is a process. 
We work with \emph{configurations} such as
\begin{equation}\label{eq:quantum-output}
\cnfig{\qstore{q,r}{\frac{1}{\sqrt{d}}\sum_{j=0}^{d-1}\ket{j}_{q}\otimes\ket{j}_{r}}}{q}{\outp{c}{q}\sep P}.
\end{equation}

This configuration means that the global quantum state consists of two
qudits, $q$ and $r$, in the specified state; that the process term under consideration has access to qudit $q$ but not to qudit $r$ ; and that the process itself is $\outp{c}{q}\sep P$. Now consider a configuration with the same quantum state but a different process term:
\[
\cnfig{\qstore{q,r}{\frac{1}{\sqrt{d}}\sum_{j=0}^{d-1}\ket{j}_{q}\otimes\ket{j}_{r}}}{r}{\outp{b}{r}\sep Q}.
\]
The parallel composition of these configurations is the following:
\[
\cnfig{\qstore{q,r}{\frac{1}{\sqrt{d}}\sum_{j=0}^{d-1}\ket{j}_{q}\otimes\ket{j}_{r}}}{q,r}{\outp{c}{q}\sep P
\parallel \outp{b}{r}\sep Q}
\]
where the quantum state is still the same.

The semantics of CQP consists of labelled transitions between
configurations, which are defined in a similar way to classical
process calculus. For example, configuration (\ref{eq:quantum-output})
has the transition
\[
\ptrns{\cnfig{\qstore{q,r}{\frac{1}{\sqrt{d}}\sum_{j=0}^{d-1}\ket{j}_{q}\otimes\ket{j}_{r}}}{q}{\outp{c}{q}\sep
  P}}{\outp{c}{q}}{\cnfig{\qstore{q,r}{\frac{1}{\sqrt{d}}\sum_{j=0}^{d-1}\ket{j}_{q}\otimes\ket{j}_{r}}}{\emptyset}{P}}.
\]
The quantum state is not changed by this transition, but because qudit $q$ is output, the continuation process $P$ no longer has access to it; the final configuration has an empty list of owned qudits. The labelled transitions were defined specifically for qubit systems and this is for the first time that the labelled transition rules have been defined specifically for higher dimensional systems (qudits), which is the focus of the present paper. 

According to the original reduction semantics of CQP, we get a probability distribution over configurations after a measurement. The next step then reduces probabilistically to one particular configuration. But in order to prove that the equivalence of CQP for qubits, between the processes is \emph{congruent}, we had to include a more sophisticated analysis of measurement in the semantics called the \emph{mixed configurations}.  Here we extend the definition of \emph{mixed configurations} with respect to qudits. We define a \emph{mixed configuration} as a weighted distribution over pure configurations.

\begin{definition}
A mixed configuration is a weighted distribution, written as 
\[
\begin{array}{l}
\Dist{i \in I}{g_i} ( \qstore{q}{\ket{\psi_{i}}} ; \omega; \ltrm{x}{P}{\tilde{v_{i}}} ) 
\end{array}
\]
with weights $g_{i}$ where $\sum_{i\in I}g_{i} = 1$ and for each $i \in I, 0 < g_{i} \le 1$ and $\ket{\psi_{i}} \in \mathbb{H}^{2|\tilde{q}|}$ and $|\tilde{v_{i}}| = |\tilde{x}|$.
\end{definition}

The operator $\oplus$ represents a distribution over the index set $I$ with weights $g_{i}$. The process term is replaced by the expression $\lambda\tilde{x}.P;\tilde{v_{i}}$ which indicates that in each component the variables $\tilde{x}$, appearing in $P$ as placeholders, should be substituted for the values $\tilde{v_{i}}$. A pure configuration can be considered as a mixed configurations with a single component.

If the observer does not get the result of a quantum measurement then we say that the system is in a mixed state. As the measurement result occurs with the process term, we need to write the configuration as a mixture which includes the mixed quantum state and the process term. Let us consider a few examples.

\begin{example}
  \label{ex:cfg_measurement}
  $
  ( \qstore{q}{\sum_{l=0}^{d-1}\alpha_{l}\ket{l}}; q; \outp{c}{\measure{q}}.P) \transition{\tau} \Dist{i \in \{0,1,..,d-1\}}{\ms{\alpha_i}} ( \qstore{q}{\ket{i}} ; q; \ltrm{x}{\outp{c}{x}.P}{i} )
  $.
\end{example}

The example illustrates a transition which represents the effect of a measurement, within a process which is going to output the result of a measurement. But the output is not part  of the transition and hence we say that this is a $\tau$ transition and the process term on the right still contains
$\outp{c}{}$. The configuration on the left is a \emph{pure configuration}, as described before. On the right we have a \emph{mixed configuration} in which the $\oplus$ ranges over the
possible outcomes of the measurement and the $\ms{\alpha_i}$ are the weights of
the components in the mixture. The quantum state
$\qstore{q}{\ket{i}}$ corresponds to the measurement outcome. The
expression $\ltrmshort{x}{\outp{c}{x}.P}$ is not a $\lambda$-calculus
function, but represents the fact that the components of the mixed
configuration have the same process structure and differ only in the
values corresponding to measurement outcomes. The final term in the
configuration, $i$, shows how the abstracted variable $x$ should be
instantiated in each component. Thus the $\lambda x$ represents a term
into which expressions may be substituted, which is the reason for the
$\lambda$ notation. So the mixed configuration
is essentially an abbreviation of
\[
\begin{array}{rcl}
\ms{\alpha_0} ( \qstore{q}{\ket{0}} ; q; \outp{c}{0}.P\subst{0}{x}) \oplus \ms{\alpha_1} ( \qstore{q}{\ket{1}}; q; \outp{c}{1}.P\subst{1}{x}\dots \\ \oplus \ms{\alpha_{d-1}} ( \qstore{q}{\ket{d-1}}; q; \outp{c}{d-1}.P\subst{d-1}{x})
\end{array}
\]

If a measurement result is output then the observer would know which of the possible states the system is in. This is represented by probabilistic branching, where we say that system to be in one branch or the other and it is no longer a mixture of components depending on the dimension $d$. 

\begin{example}\label{ex:cfg_output}
\[
    \Dist{i \in \Omega}{\ms{\alpha}_i} ( \qstore{q}{\ket{i}} ; q; \ltrm{x}{\outp{c}{x}.P}{i} ) \transition{\outp{c}{\Omega}}  \Prob{i \in \Omega}{\ms{\alpha_i}} ( \qstore{q}{\ket{i}} ; q; \ltrm{x}{P}{i} ) 
    \ptrans{\ms{\alpha_0}}  ( \qstore{q}{\ket{0}} ; q; \ltrm{x}{P}{0} ) 
\]  

\end{example}
Example~\ref{ex:cfg_output} shows the effect of the output from the final configuration of Example~\ref{ex:cfg_measurement}. The output transition produces the intermediate configuration, which is a
probability distribution over pure configurations (which is represented as  the change from $\oplus$ to
$\boxplus$). Because it comes from a mixed configuration, the output transition contains a \emph{set} of possible values. From the intermediate configuration there are probabilistic transitions and the number of transitions depends on the dimension $d$, of which one is shown ($\ptrans{\ms{\alpha_0}}$). 
Here $\Omega$ is a set of values given by \{0,1,..,d-1\}. Measurement outcomes may be communicated between processes without creating a probability distribution. In these cases an observer must
still consider the system to be in a mixed configuration. 

\begin{example}
  \label{ex:cfg_communication}
\[
    \Dist{i \in \Omega}{g_i} ( \qstore{q}{\ket{i}} ; q; \ltrm{x}{(\outp{c}{x}.P \parcomp \inp{c}{y}.Q)}{i} ) 
    \transition{\tau} \Dist{i \in \Omega}{g_i} (
    \qstore{q}{\ket{i}} ; q; \ltrm{x}{(P \parcomp Q\subst{x}{y})}{i} )
\]
\end{example}

In Example~\ref{ex:cfg_communication} there is a mixed configuration on
the left, with arbitrary weights $g_i$, which we imagine to have been
produced by a measurement.  If we now have a receiver for the
output, there will be no difference in process $Q$ between the $d$
components of the mixed configuration. However, after communication, the different values for $x$ have been propagated to $Q$, so we include $Q$ in the abstraction.

\subsubsection{ Expression Transition Relations for qudits}
In this section, we provide and discuss the necessary transition rules which has been modified for qudits. The transition relations $\transition{}_{v}$ for evaluating values and $\transition{}_{e}$ for evaluating expressions are defined by the rules in Figure~\ref{fig:trans_exp}. Rules $\Rplus$ and $\Rtrans$ deal with the evaluation of terms that result in values. $\Rplus$ introduces a variable $x$ as a placeholder for the value $w$. The placeholder is important as when we consider mixed expression configurations in $\Rcontext$, there may be a different value resulting from each component.

The transition relations for qudit are similar to that for qubits given in \cite{DavidsonThesis}. The difference is in the rule $\Rmeasure$. Since qudits are $d$-level quantum systems, we need to consider the number of dimensions as $d$, compared to qubits which is two. $\Rmeasure$ is a measurement rule which produces a mixed configuration in which each component corresponds to a specific outcome $m$. The variable $x$ is to maintain a constant expression term across all components, while the measurement value $m$ is different for each component. Applying a unitary operator always results in the value unit, hence $\Rtrans$ does not introduce a new variable.

The rule $\Rcontext$ has two main purposes, it is used for the evaluation of expressions in an expression context $E$ and is also used for the evaluation. The evaluation of a mixed expression configuration configuration  $\Dist{i \in I}{h_i}( \sigma_{i}; \omega; \lambda \tilde{y}.E[e];\tilde{u_{i}})$ is determined by the evaluation of each component.
\begin{figure*}
  \begin{gather*}
    \tag{\Rplus}
    ( \qstore{\vec{q}}{\ket{\psi}} ; \omega; u + v) \transitionv ( \qstore{\vec{q}}{\ket{\psi}} ; \omega; \ltrm{x}{x}{w}) \quad \textrm{where $w = u + v$} \\
    \tag{\Rmeasure}
    \begin{array}{r}
      ( \qstore{q_0,\dots,q_{n-1}}{\alpha_0 \ket{\phi_0} + \dots + \alpha_{d^n-1} \ket{ \phi_{d^n-1}}}; \omega; \measure{ q_0,\dots, q_{r-1} } ) \transitionv \\
      \Dist{0 \le m < d^r}{g_m} ( \qstore{q_0, \dots, q_{n-1}}{\frac{\alpha_{l_m}}{\sqrt{g_m}} \ket{ \phi_{l_m}} + \dots + \frac{ \alpha_{u_m}}{\sqrt{g_m}} \ket{ \phi_{u_m}}}; \omega; \ltrm{x}{x}{m} )
    \end{array} \\
    \textrm{where } l_m = d^{n-r}m, u_m = d^{n-r}(m+1)-1, g_m = |\alpha_{l_m}|^2 + \dots + |\alpha_{u_m}|^2 \\
    \tag{\Rtrans}
    ( \qstore{q_0,\dots,q_{n-1}}{\ket{\phi}} ; \omega; q_0,\dots,q_{r-1}\trans{U^m} ) \transitionv \hspace{20mm} \\
    \hspace{30mm} (\qstore{q_0,\dots,q_{n-1}}{(U^m \otimes I_{n-r}) \ket{\phi}}; \omega; \unit; \cdot ) \\
    \tag{\Rcontext}
    \begin{prooftree}
      \forall i \in I. ( \qstore{\vec{q}}{\ket{\psi_i}} ; \omega; e\{\vec{u}_i/\vec{y}\}) \transitionv \Dist{j \in J_i}{g_{ij}} ( \qstore{\vec{q}}{\ket{\psi_{ij}}} ; \omega; \ltrm{\vec{x}}{e'\{\vec{u}_i/\vec{y}\}}{\vec{v}_{ij}} )
      \justifies
      \Dist{i \in I}{h_i} ( \qstore{\vec{q}}{\ket{\psi_i}} ; \omega; \ltrm{\vec{y}}{E[e]}{\vec{u}_i}) \transitione \Dist{\substack{i \in I\\j \in J_i}}{h_ig_{ij}} ( \qstore{\vec{q}}{\ket{\psi_{ij}}} ; \omega; \ltrm{\vec{y}\vec{x}}{E[e']}{\vec{u}_i,\vec{v}_{ij}})
    \end{prooftree}
  \end{gather*}
  \caption{Transition rules for values and expressions.}
\label{fig:trans_exp}
\end{figure*}
With the new set of transition rules, we will be able to describe the execution of the higher dimensional quantum protocols namely teleportation and superdense coding in the next sections.

\subsection{Execution of Teleportation}

We present the interesting steps in one possible execution of $\pteleport$. The semantics of CQP is not deterministic and hence the transitions can proceed in a different order. The order shown here is chosen for presentation convenience. Consider a qudit to be teleported is given by the quantum state $\ket{\psi} = \sum_{l=0}^{d-1}\alpha_{l}\ket{l}$. The initial configuration is $((\tilde{r}x = \sum_{l=0}^{d-1}\alpha_{l}\ket{l}_{x});\emptyset;\pteleport)$. In the first few steps, the system executes  $\Qdit$ terms, the Hadamard operation and the CNOT Right-Shift ($R_{C}$), constructing the global quantum state:
\[
(\tilde{r}pq_{1}q_{2} = \sum_{l=0}^{d-1}\alpha_{l}\ket{l}_{x} \otimes \frac{1}{\sqrt{d}}\sum_{k=0}^{d-1}\ket{k}_{q_{2}}\otimes \ket{k}_{q_{1}});q_{1},q_{2};(\new e)(\palice\action{q_{2}/z} \parallel \pbob\action{q_{1}/y}))
\]
$\palice$ receives the qudit $x$, in state $\psi$, from the environment, through the input transition $^{\inp{c}{x}}_{\longrightarrow}$, which gives us the $3$ qudit state. After some $\tau$ transitions corresponding to $\palice$$'s$ Hadamard and CNOT Left-Shift ($L_{C}$) operations, we have:
\[
(\tilde{r}xq_{1}q_{2} = \ket{\Phi_{2}});q_{1},q_{2},p;(\new e)(\outp{e}{\measure q_{2},\measure x}\sep\nil \parallel \pbob\action{q_{1}/y}))
\]
where $ \ket{\Phi_{2}} = \frac{1}{d}\sum_{l,j,k=0}^{d-1}\omega^{-lj}\alpha_{l}\ket{j}_{x} \otimes \ket{k \ominus l}_{q_{2}}\otimes \ket{k}_{q_{1}}$.
$\palice$ does the measurement of her qudits in the \emph{standard basis} and the results are communicated to $\pbob$ via channel $e$. Since the communication is internal within the system, this produces a mixed configuration which is given as:
\[
 \oplus_{j \in \Omega, s \in \Omega}((\tilde{r}xq_{1}q_{2} = \ket{\Psi_{js}});q_{1},q_{2},x;\lambda_{M_{1},M_{2}}\sep(\new e)(\outp{e}{M_{1},M_{2}}\sep\nil \parallel \pbob\action{q_{1}/y});j,s)
 \]
where  $\ket{\Psi_{js}} = \frac{1}{d^{2}}\sum_{j,s=0}^{d-1}\ket{j}_{x}\ket{s}_{q_{2}}\sum_{l=0}^{d-1}\omega^{-lj}\alpha_{l} \ket{l \oplus s}_{q_{1}}$. Depending on the classical values ($M_{1}$ and $M_{2}$) $\pbob$ does his unitary operations on his qudit $q_{1}$ to get the same state of the qudit $x$ which $\palice$ possesses. The qudit is then output through channel $d$.
\[
\oplus_{j \in \Omega, s \in \Omega}((\tilde{r}pq_{1}q_{2} = \ket{\Psi'_{js}});q_{2},p;\lambda_{M_{1},M_{2}}\sep\nil;j,s)
\]
where $\ket{\Psi'_{js}} = \frac{1}{d^{2}}\sum_{l=0}^{d-1}\alpha_{l}\ket{l}_{q_{1}}$.

%% file: SDC.tex
\section{Superdense Coding Protocol for Qudits}
\label{sec:SDC}
\label{sec-SDC}

In this section, we will describe the superdense coding protocol with respect to qudits. This protocol is considered the opposite of teleportation, where two values of classical information are communicated by exchanging a single qudit. Superdense coding also involves two users sharing a pair of entangled qudits. The quantum circuit for this protocol is given in Figure~\ref{fig:sdc}. The goal is to transmit some classical information from one user (Alice) to another (Bob). Like the previous protocol, this also begins with the preparation of an entangled pair. Alice is in possession of the first qudit, while Bob has possession of the second qudit.  By sending the single qudit in her possession to Bob, it turns out Alice can communicate two classical values (ranging from 0 to $d-1$) to Bob ,where $d$ is the dimension of the system. The CQP definition of the system is  given below:
\[
\begin{array}{ll}
\multicolumn{2}{l}{\psdc = (\Qdit: q_{1},q_{2})(\action{q_{1}\trans\gH}\sep\action{q_{1},q_{2}\trans\gRC}\sep(\new e)(\palice(c,e) \parallel \pbob(e,d)))}\\
\\ \multicolumn{2}{l}{\palice(\typed{c}{\chant{\Val,\Val}},\typed{e}{\chant{\Qdit}}) = \inp{c}{\typed{a}{\Val},\typed{b}{\Val}}\sep\action{q_{1}\trans\gX^{b}}\sep\action{q_{1}\trans\gZ^{a}}\sep\outp{e}{q_{1}}\sep\nil} \\
\\ \multicolumn{2}{l}{\pbob(\typed{e}{\chant{\Qdit}},\typed{d}{\chant{\bit,\bit}}) = \inp{e}{\typed{q_{1}}{\Qdit}}\sep\action{q_{1},q_{2}\trans\gLC}\sep\action{q_{1}\trans\gH}\sep\outp{d}{\measure q_{1},\measure q_{2}}\sep\nil} \\
\end{array}
\]
This CQP model, unlike the circuit model (Figure~\ref{fig:sdc}), is able to clearly describe the actions of the two users using the processes $\palice$ and $\pbob$. $\palice$ takes one qudit ($q_{1}$) of the pair and $\pbob$ takes the other ($q_{2}$). The classical values to be transmitted are labelled \emph{a} and \emph{b}. When $\palice$ is ready to send, she applies a combination of the $X$ and $Z$ operators to qudit $q_{1}$ depending on the values $a$ and $b$.\\
\\After $\palice$ has done her encoding, she send her single qudit to $\pbob$. Now that $\pbob$ has both qudits, he can determine which encoding $\palice$ used, and therefore the corresponding values $a$ and $b$. First, he applies a CNOT Left shift operator to qudits $q_{1}$ and $q_{2}$, followed by the Hadamard operator applied to $q_{1}$. He then measures both of these qudits to reveal the respective values. Since the state he measures is not a superposition, the outcome will be certain.    
\subsection{Execution of SDC}
 \begin{figure}
 \Qcircuit @C=1em @R = 2.0em {
     & & & & & &  & & & & & & & &\lstick{a} & \cw & \cw & \cw & \cw & \control \cw \cwx[2] & \\
     & & & & & &  & & & & & & & &\lstick{b} & \cw & \cw & \control \cw \cwx[1] & \\
    & & & & & & & & & & & &\lstick{y = \ket{0}}       &     \gate{H} & \ctrl{1} & \qw  & \qw & \gate{X} & \qw & \gate{Z} & \qw & \ctrl{1}  & \gate{H} & \qw & \meter & \cw & \rstick{a}\\   
     & & & & & & & & & & & &\lstick{z = \ket{0}}      &     \qw &  \gate{R_C} & \qw & \qw & \qw & \qw & \qw & \qw & \gate{L_C}  & \qw  & \qw & \meter & \cw & \rstick{b}
}
\caption{\label{fig:sdc}Superdense Coding Protocol}
\end{figure} 
In this section, we show the step-by-step execution of the superdense coding protocol, in order to illustrate the operational semantics. Teleportation can also be executed in a similar way according to the transition rules.

Consider an arbitrary quantum state $\tilde{r} = \ket{\psi}$. Let $s = (\tilde{r} = \ket{\psi};\emptyset;\psdc)$, then the execution of superdense coding is as follows.
 \[
 \begin{array}{l}
 s {\weaktrans{\tau}}((\tilde{r}q_{1}q_{2} = \ket{\psi_{1}});q_{1},q_{2};(\new e)(\palice(c,e) \parallel \pbob(e,d))\\

{\transition{\inp{c}{a,b}}}((\tilde{r}q_{1}q_{2} = \ket{\psi_{1}});q_{1},q_{2};(\new e)(\action{q_{1}\trans\gX^{b}}\sep\action{q_{1}\trans\gZ^{a}}\sep\outp{e}{q_{1}}\sep\nil \parallel \pbob(e,d)))\\

\\{\transition{\tau}}((\tilde{r}q_{1}q_{2} = \ket{\psi_{2}});q_{1},q_{2};(\new e)(\action{q_{1}\trans\gZ^{a}}\sep\outp{e}{q_{1}}\sep\nil \parallel \pbob(e,d)))\\

\\{\transition{\tau}}((\tilde{r}q_{1}q_{2} = \ket{\psi_{3}});q_{1},q_{2};(\new e)(\outp{e}{q_{1}}\sep\nil \parallel \pbob(e,d)))\\

\\{\transition{\tau}}((\tilde{r}q_{1}q_{2} = \ket{\psi_{3}});q_{1},q_{2};(\new e)(\action{q_{1},q_{2}\trans\gLC}\sep\action{q_{1}\trans\gH}\sep\outp{d}{\measure q_{1},\measure q_{2}}\sep\nil))\\

\\{\transition{\tau}}((\tilde{r}q_{1}q_{2} = \ket{\psi_{4}});q_{1},q_{2};(\new e)(\action{q_{1}\trans\gH}\sep\outp{d}{\measure q_{1},\measure q_{2}}\sep\nil))\\

\\{\transition{\tau}}((\tilde{r}q_{1}q_{2} = \ket{\psi_{5}});q_{1},q_{2};(\new e)(\outp{d}{\measure q_{1},\measure q_{2}}\sep\nil))\\

\\{\transition{\outp{d}{a,b}}}((\tilde{r}q_{1}q_{2} = \ket{\psi_{6}});q_{1},q_{2};\nil)\\
\\ where
 \end{array}
 \]
 \[
 \begin{array}{l}
 \ket{\psi_{1}} = \frac{1}{\sqrt{d}}\sum_{j=0}^{d-1}\ket{j}_{q_{1}} \otimes \ket{j}_{q_{2}}\\
 \\  \ket{\psi_{2}} = \frac{1}{\sqrt{d}}\sum_{j=0}^{d-1}\ket{j \oplus b}_{q_{1}} \otimes \ket{j}_{q_{2}}\\
 \\  \ket{\psi_{3}} = \frac{1}{\sqrt{d}}\sum_{j=0}^{d-1}\omega ^{a(j \oplus b)}\ket{j \oplus b}_{q_{1}} \otimes \ket{j}_{q_{2}}\\
  \\  \ket{\psi_{4}} = \frac{1}{\sqrt{d}}\sum_{j=0}^{d-1}\omega ^{a(j \oplus b)}\ket{j \oplus b}_{q_{1}} \otimes \ket{j \ominus (j \oplus b)}_{q_{2}} =  \frac{1}{\sqrt{d}}\sum_{k=0}^{d-1}\omega ^{ak}\ket{k}_{q_{1}} \otimes \ket{-b}_{q_{2}}\\
   \\  \ket{\psi_{5}} = \ket{\psi_{6}} = \ket{a}_{q_{1}}\otimes\ket{-b}_{q_{2}}\\
 \end{array}
 \]

%% file: conclusion.tex
\section{Conclusion and Future Work}
\label{sec-conclusion}
\label{sec:conclusion}

We have explained the use of the quantum process calculus CQP, and extended it to 
model $d$-dimensional quantum systems. With the help of the generalised quantum gates and generalised Bell states, we have applied CQP to describe two quantum protocols, namely teleportation and superdense coding for higher dimensional systems. The next  task would be to extend the CQP definitions with respect to orbital angular momentum of light and this would provide a platform to express or model in CQP the real quantum information processing systems which are used in quantum optical experiments. \\
\\Quantum protocols can be represented by quantum circuits but it does not provide the full description of the protocol. Although the circuit model defines the operation involved in the protocol, it does not indicate the number of users involved in the protocol. But with process calculus, we have a formal definition of the users or the processes involved in the protocol. In addition to a clear description of the actions of the processes, we also have an indication of the way in which the processes communicate.
Quantum process calculus provides a systematic methodology for verification of quantum systems and in the previous work of CQP, the behavioural equivalence of CQP \cite{DavidsonThesis} is defined with respect to qubits and also the equivalence is proved to be a congruence. Analysis and verification of quantum protocols were done with respect to qubits. Although we have not yet defined a theory of behavioural equivalence for qudits, we believe it should be a straightforward task. 
The fact that CQP can also express classical behaviour, means that we have a uniform framework in which to analyze classical and quantum computation and communication. The long-term goal is to develop software for automate analysis of CQP models, including both qubits and qudits, following established work in classical process calculus.\\